\documentclass[aps,prd,onecolumn,superscriptaddress,amsfont,graphicx,nofootinbib,preprintnumbers]{revtex4}%
\usepackage{color,graphicx,epsfig}
\usepackage{ifpdf}
\usepackage{amsmath}
\usepackage{bm}
\usepackage{color}
\usepackage[english]{babel}
\usepackage{graphicx}%
\usepackage{amsfonts}%
\usepackage{amssymb}
\usepackage{braket}
\usepackage{hyperref}

\usepackage{array}
\newcolumntype{L}[1]{>{\raggedright\let\newline\\\arraybackslash\hspace{0pt}}m{#1}}
\newcolumntype{C}[1]{>{\centering\let\newline\\\arraybackslash\hspace{0pt}}m{#1}}
\newcolumntype{R}[1]{>{\raggedleft\let\newline\\\arraybackslash\hspace{0pt}}m{#1}}

\definecolor{nicered}{rgb}{0.7,0.1,0.1}
\definecolor{nicegreen}{rgb}{0.1,0.5,0.1}
\hypersetup{colorlinks,citecolor= nicegreen,linkcolor= nicered}

\newcommand{\beq}{\begin{equation}}
\newcommand{\eeq}{\end{equation}}
\newcommand{\bea}{\begin{eqnarray}}
\newcommand{\eea}{\end{eqnarray}}

\begin{document}
\topmargin -1.0cm
\oddsidemargin -0.8cm
\evensidemargin -0.8cm
\begin{flushright}
ICAS 030/17 \\
\end{flushright}

\def\LjubljanaFMF{Faculty of Mathematics and Physics, University of Ljubljana,\\
 Jadranska 19, 1000 Ljubljana, Slovenia }
\def\LjubljanaIJS{Jo\v zef Stefan Institute, Jamova 39, 1000 Ljubljana, Slovenia}

\def\BuenosAires{International Center for Advanced Studies (ICAS) and CONICET, UNSAM, Campus Miguelete\\
25 de Mayo y Francia, (1650) Buenos Aires, Argentina}

\def\CONICET{Centro At\'omico Bariloche, Instituto Balseiro and CONICET\\
 Av. Bustillo 9500, 8400, S. C. de Bariloche, Argentina}

\title{Measuring $|V_{td}|$ at LHC}

\author{Ezequiel Alvarez} 
\email[Electronic address: ]{sequi@df.uba.ar} 
\affiliation{\BuenosAires}

\author{Leandro Da Rold} 
\email[Electronic address: ]{daroldl@cab.cnea.gov.ar} 
\affiliation{\CONICET}

\author{Mariel Estevez} 
\email[Electronic address: ]{mestevez@unsam.edu.ar} 
\affiliation{\BuenosAires}

\author{Jernej F.\ Kamenik} 
\email[Electronic address: ]{jernej.kamenik@ijs.si} 
\affiliation{\LjubljanaIJS}
\affiliation{\LjubljanaFMF}

\preprint{}

\begin{abstract}
We propose a direct measurement of the CKM element $V_{td}$ at the LHC. Taking profit of the imbalance between $d$ and $\bar d$ quark content in the proton, we show that a non-zero $V_{td}$ induces a charge asymmetry in the $tW$ associated production. The main backgrounds to this process, $t\bar t$ production, and $tW$ associated production mediated by $V_{tb}$, give charge symmetric contributions at leading order in QCD.  Therefore, using specific kinematic features of the signal, we construct a charge asymmetry in the di-lepton final state which, due also to a reduction of systematic uncertainties in the asymmetry, is potentially sensitive to $V_{td}$ suppressed effects.  In particular, using signal and background simulations up to detector level, we show that this new observable could improve the current direct upper bound on $|V_{td}|$ already with existing LHC data.  We also project that $|V_{td}|$ values down to $\sim 10$ times the Standard Model prediction could be probed in the high luminosity phase of the LHC.
\end{abstract}

\maketitle

%
\section{Introduction}
%

The entries in the CKM matrix governing flavor transitions among quarks are fundamental parameters of the standard model (SM). As such they warrant intense experimental scrutiny. Currently the first two rows of the CKM are already being probed directly with ever improving precision using decays of nuclei, kaons, charmed mesons and B-hadrons~\cite{Patrignani:2016xqp}. On the other hand, few direct experimental handles exist on the third row of the CKM. The SM predictions for $V_{tq}$ matrix elements, with $q=d,s,b$ are currently derived from CKM unitarity considerations, as well as measurements of radiative decays and oscillations of B-mesons, where $V_{tq}$ enter in loops involving virtual top quarks. A recent global CKM fit yields~\cite{Charles:2015gya}
\beq
|V^{\rm SM}_{tb}| = 1-8.81^{+0.12}_{-0.24} \times 10^{-3}\,, \quad |V^{\rm SM}_{ts}| = 41.08^{+3.0}_{-5.7} \times 10^{-3}\,, \quad |V^{\rm SM}_{td}| = 8.575^{+0.076}_{-0.098} \times 10^{-3}\,.
\label{vtd}
\eeq
The aim is thus to confront these indirect $|V_{tq}|$ determinations using direct measurements of processes involving on-shell top quarks. The LHC, as currently the only top-quark production machine, can probe $V_{tq}$ directly by studying top production as well as decays. In particular, measurements of b-jet fractions in top decays $t \to W j$ currently put a bound on~\cite{Khachatryan:2014nda}
\beq
\frac{\mathcal B(t\to b W)}{\sum_{q=d,s,b} \mathcal B(t \to q W)} > 0.955~ @ ~95\% ~\rm C.L.\,,
\eeq
which can be interpreted as $\sqrt{|V_{td}|^2+|V_{ts}|^2} < 0.217 |V_{tb}|$\,. In addition, precise measurements of $t$-channel single top production and its charge asymmetry at the LHC~\cite{Aaboud:2016ymp,Aaboud:2017pdi, Sirunyan:2016cdg} when compared with the accurate theoretical predictions~\cite{Campbell:2009ss,Kant:2014oha,Brucherseifer:2014ama} can be interpreted as measurements of $|V_{tq}|$. A recent ATLAS analysis neglecting $|V_{td,ts}|$ effects yields
$
|V_{tb}| = 1.07 \pm 0.09
$~\cite{Aaboud:2016ymp}\,. 
While such measurements are in principle also able to probe the $|V_{ts}|$ and $|V_{td}|$ matrix elements~\cite{Lacker:2012ek}, they are intrinsically limited by the overwhelming backgrounds and associated statistical and systematic uncertainties. Consequently, especially in the case of $|V_{td}|$ they are not expected to come even close to the magnitude of the SM predictions. 

In the present work we outline an experimental strategy to probe the $|V_{td}|$ matrix element directly at the LHC using single top production associated with a $W$ boson ($p p \to t W$)\,.\footnote{For a previous study of possible NP effects in this mode see Ref.~\cite{Tait:2000sh}.} Our proposal exploits the production cross-section enhancement as well as boosts of the top quarks coming from initial state valence $d$-partons. In addition, and contrary to $t$-channel single top production, the main backgrounds have vanishing or very small charge asymmetries. This opens a path towards direct $|V_{td}|$ determination at the (HL)LHC within an order of magnitude of the SM prediction.

The remainder of the paper is structured as follows. In Sec.~\ref{sec:Vtd} we review the main effects of non-vanishing $V_{td}$ on top quark production processes at hadron colliders. In Sec.~\ref{sec:Assym} we focus on the charge asymmetry in $Wt$ production analyzing the dominant backgrounds and proposing an analysis strategy to reduce these while preserving most of the signal. The main results of work are presented in Sec.~\ref{sec:results} with conclusions drawn in Sec.~\ref{sec:conclusions}.

\begin{figure}[t]
\centering
\includegraphics[width=0.5\textwidth]{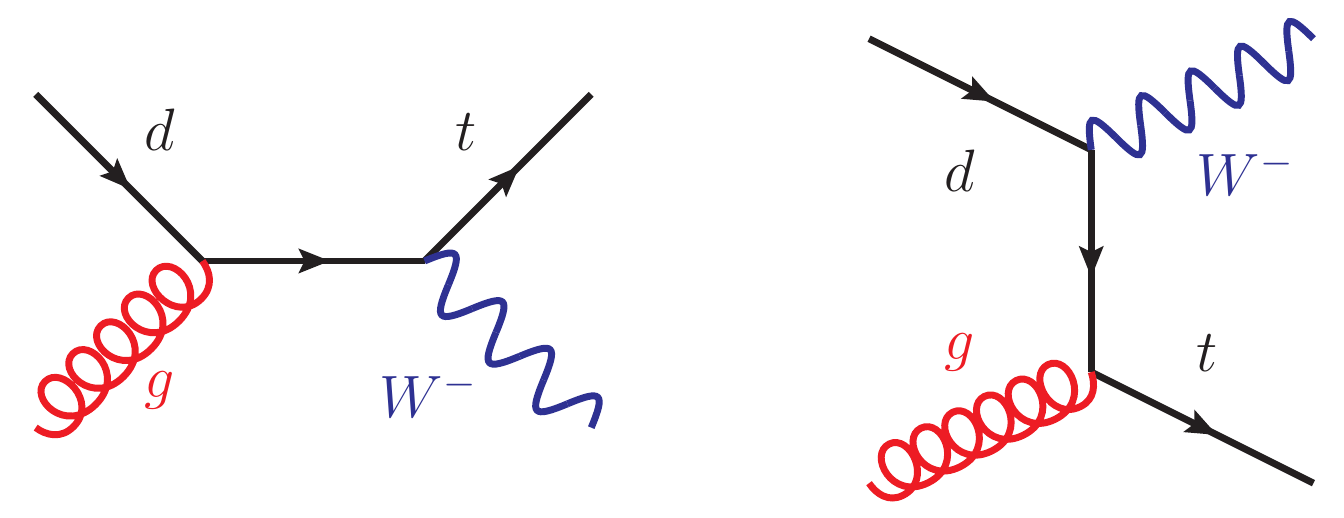}
\caption{Feynman diagrams contributing to $pp \to tW$ production proportional to $V_{td}$ at LO in QCD.} 
\label{signal}
\end{figure}

\section{LHC processes and observables sensitive to $V_{td}$}
\label{sec:Vtd}

Production of the top quark from an initial down quark in the proton is highly suppressed because of the expected smallness of $|V_{td}|\sim {\cal O}(10^{-2})$. Any relevant process mediated via the $tWd$ coupling is expected to produce a tiny signal, and thus would be difficult to measure directly at the LHC, both because of the small expected statistics and large backgrounds, but especially because of systematic uncertainties inherent to a hadronic machine.  Therefore, observability of $V_{td}$-sensitive processes at the LHC is closely related to the capability of finding associated observables with reduced experimental systematic uncertainties.  Common approaches to taming systematics include data-driven methods and asymmetries. Since the $d$-quark is a valence constituent of the proton, its imbalance with the $\bar d$-quark together with the charge self-tagging of leptonically decaying $W$ bosons and top quarks motivates to explore $V_{td}$-sensitive observables in the form of charge asymmetries. In the following we will parameterize eventual departures from SM in $V_{td}$ through the ratio
\begin{equation}
r \equiv \left| \frac{V_{td}}{V_{td}^{\rm SM}}\right|\,,
\end{equation}
in order to classify processes according to their leading power in $r$.

As a first process, we discuss the $tW$ associated production mediated by the partonic process $g d \to t W^-$ (see Fig.~\ref{signal} for the relevant leading order (LO) Feynman diagrams), whose cross-section is proportional to $|V_{td}|^2$ (and thus $r^2$). At LO in the SM $\sigma(tW^-) = 20$ fb \cite{Alwall:2014hca} at the 13 TeV LHC, while for the CP-conjugate final state $\sigma(\bar t W^+)=6$ fb. This process is interesting both because of its sizeable charge asymmetry but also because its kinematics predicts a characteristic angular distribution.  In fact, the dominant diagram has a virtual top quark exchanged in the t-channel. Because of the relatively large incoming momentum expected on average from a valence $d$-quark, it consequently prefers a forward $W^-$ in the lab frame.  This special feature allows to consider the interesting di-lepton final state, which permits a relatively clean search strategy.  In fact, this forward preference of the $W^-$ is translated in having a preferably forward $\ell^-$ in signal events.  The two main backgrounds to this $\ell^+ \ell^- b$ final state would be the di-leptonic $t\bar t$ production (missing one of the $b$-jets from top decays) and $tW$ associated production proportional to $|V_{tb}|^2$ (also in the leptonic decay channel of both the $t$ and the $W$)\,. Since this process features a collinear $b$ coming from initial state gluon splitting (either re-summed as in the five-flavor PDF scheme, or explicit as in the four-flavor PDF scheme), in the following we denote is as $tW(b)$ production. Importantly, both backgrounds have very small charge asymmetries, and we expect a charge asymmetry constructed with this final state to exhibit promising sensitivity to $V_{td}$.  Other reducible backgrounds and their relevance are discussed in more detail in the next section.

\begin{figure}[t!]
\centering
\includegraphics[width=0.5\textwidth]{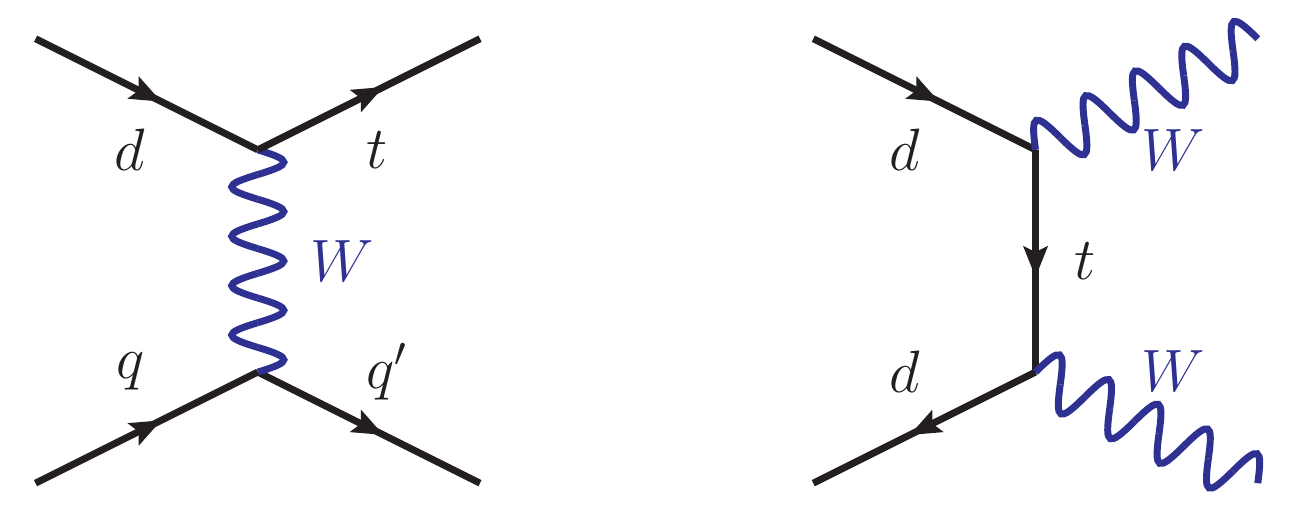}
\caption{Examples of Feynman diagrams contributing to further $V_{td}$-sensitive processes at the LHC.}
\label{feynmanotros}
 \end{figure}

{Another potentially important  $V_{td}$-sensitive process is $pp \to t j$, where partonic processes such as $d q \rightarrow t q'$ and $d \bar q' \rightarrow t \bar q$ yield contributions that go as $|V_{td}|^2$, see Fig.~\ref{feynmanotros}. (Here  $q=u,c$ and $q'=d,s,b$.) This process has a contribution where both initial quarks are valence quarks, $d u \to t d$, and therefore it is enhanced with respect to the contribution of its CP conjugate, producing a charge asymmetry.}  However,  its main background, the $t$-channel single top production ($p p \to t j (b)$) proportional to $|V_{tb}|^2$ is also significantly charge asymmetric. The sensitivity in this channel is thus limited by both the theoretical knowledge of its SM prediction and experimental systematics in its measurement, and we do not pursue it further.

{So far we have discussed signals whose cross-sections are proportional to $|V_{td}|^2$ ($r^2$). One could however also consider (tree-level) processes contributing at higher orders in $V_{td}$. For instance, the contribution to $WW$ production coming from $d \bar{d} \rightarrow W^{+} W^{-}$ with a top quark exchanged in the $t$-channel is asymmetric in the angular distribution of the final state particles and has a term proportional to $|V_{td}|^{4}$ ($r^4$). (See Fig.~\ref{feynmanotros} for the relevant LO Feynman diagram on the right-hand side.)}
However, one of the main backgrounds to this signal would be the $W^{+} W^{-}$ production mediated through the $t$-channel exchange of an up- or charm-quark, which has a similar asymmetry as the signal.  Again in this case the usefulness of the asymmetry is reduced because one would need to compare it to a non-negligible reference number of the background.  However, there are two features of this process that should be mentioned and which may deserve further exploration.  One is that its cross-section has a term proportional to $r^4$, and although this contribution is suppressed by $V_{td}^4$, it is doubly enhanced compared to previous ones for $r>1$.
The second feature is that this process would affect both the angular and invariant mass distributions of the $WW$ final state, and a sensitive observable could be constructed using side-band fitting.

There exist further $V_{td}$-sensitive processes which we do not discuss here. Instead in the remainder of the paper, we focus on $pp \to tW$ and study the prospects of measuring or constraining $V_{td}$ through a suitably defined charge asymmetry in the di-lepton final state channel.

\section{Asymmetry sensitive to $V_{td}$}
\label{sec:Assym}

In this section we explore potential direct experimental sensitivity to $V_{td}$ through the process $pp \to tW$ at the LHC.  As discussed in the previous section, the kinematics of the signal offers the opportunity to distinguish the leptons coming from the $W$ and the top decay, and therefore allows for a search strategy using a charge asymmetry in the $\ell^+ \ell^-b$ final state.
In the following sub-sections we first analyze the relevant backgrounds before constructing a suitable $V_{td}$-sensitive charge asymmetry.  We end by discussing other charge asymmetries which could provide further valuable observables in other contexts.

\subsection{Backgrounds}
In the following we discuss the main backgrounds and give an estimate of their size. The actual numbers used in our results are obtained through simulations described in the next section. We first note that the ATLAS and CMS collaborations have analyzed this process in both 8TeV and 13TeV LHC data~\cite{Chatrchyan:2014tua, Aad:2015eto, Aaboud:2016lpj}. Here we roughly follow Ref.~\cite{Chatrchyan:2014tua} in the detailed characterization of the backgrounds, considering the final state with two leptons and a $b$-jet. We start by discussing the main backgrounds first.

The $t\bar t$ production in the fully leptonic channel with a missed $b$-jet leads to the same final state as the signal. Being produced by QCD interactions, this is the dominant background, with a LO cross-section of order $\sigma(t\bar t)\sim 500$~pb and at NLO $\sigma(t\bar t) \sim 680$~pb (both estimated through {\tt MadGraph5\_aMC@NLO} \cite{Alwall:2014hca}). {Requiring exactly one $b$-jet within the typical detector acceptance ($|\eta_b|<2.5$ and $p_T(b) > 20$~GeV) and no other jet with $|\eta|<5$ and $p_T>20$ GeV, leads to a considerable reduction of this background, suppressing it by a factor $\sim 10^{-2}$ at parton level}. At LO $t\bar t$ arises from gluon fusion or from $q\bar q$ with a gluon in an $s$-channel, and so is completely charge symmetric. However, having a large cross section, it suppresses any net charge asymmetry by contributing to its denominator. At NLO this background does give a small asymmetric contribution, its size depending on the specific definition of the charge asymmetry, as described in the next section.

Another important background is given by the $tW^-(\bar b)$ and $\bar tW^+(b)$ final states, without intermediate on-shell $t$ or $\bar t$.~\footnote{In a four-flavor PDF scheme, this background is dominated by gluon fusion similar to $t\bar t$, whereas in the five-flavor PDF scheme the LO partonic process is $g b \to tW$ and the final state matches the signal exactly. See e.g.~Ref.~\cite{Frixione:2008yi} for details on separating this process from $t\bar t$ production in simulations.} {The LO production cross-section is $\sigma(tW(b))\sim 28$\,pb \cite{Alwall:2014hca}. } 
At NLO this background is expected to develop a small charge asymmetry, but since its cross-section is considerably smaller than $t\bar t$, in practice we can safely neglect it.

The Drell-Yan dominated $\ell^+\ell^-j$ production, with $j$ misidentified as a $b$-jet, is in principle an important background. Since the lepton pair arises from an intermediate $Z$ or $\gamma$,  $\ell^+$ and $\ell^-$ have the same flavor. The LO cross-section is of the order $\sigma(\ell\ell j)\sim 440$~pb. The presence of the jet induces a charge asymmetric distribution. This background can be drastically reduced by demanding different flavors of the final-state  leptons at the expense of loosing half of the the signal. A significant reduction is instead obtained by demanding $m_{\ell\ell}$ larger than 25 GeV and excluding a region around the $Z$ mass, that we choose between 75 and 105 GeV. Besides, since the signal has missing energy from the undetected neutrino arising from the leptonic decay of the top and $W$, whereas there is no missing energy for this background, we demand $E_T^{\rm miss}>30$~GeV. These cuts, in addition to a rejection factor of mis-tagging the light jet as a $b$, make the final contribution of this background to the cross-section negligible.

There is also a background similar to the previous one, but with a $b/\bar b$ pair in the final state: $\ell^+\ell^-b/\bar b$. The LO cross-section is of the order $\sigma(\ell\ell b)\sim 32$~pb, but cuts in $m_{\ell\ell}$ and  $E_T^{\rm miss}$ reduce this background as in the case of $\ell^+\ell^-j$. Although in the present case there is no significant rejection-factor associated with the jet(s), the asymmetry in $\ell^+ \ell^- b$ is much smaller than the one of $\ell^+\ell^-j$, as can be expected since the $b$-quark is not a valence quark.

The $t$-channel single top production ($tj(b)$ and $\bar t j(b)$), with the $j$ misidentified as a lepton, gives a large asymmetric background: $\sigma(tj(b))\simeq 52$~pb and $\sigma(\bar tj(b))\simeq 35$~pb, at LO. A lepton mis-identification rate of the order $\sim 10^{-4}$~\cite{Alvarez:2016nrz} suppresses this background, leading to a negligible cross-section.  

Other backgrounds include $WWj$, with $j$ being either a light or heavy ($b$ or $c$) jet flavor. The first case has a sizable charge asymmetry, although also a large rejection factor. The second case has a small rejection factor, but a tiny charge asymmetry.  We have verified that both of these backgrounds are unimportant.

We note that some of the backgrounds listed above have contributions that depend on $V_{td}$, and could be enhanced for $r>1$. However, even for $r\sim{\cal O}(10)$, the size of the backgrounds does not change significantly. In particular, the $t\bar t$ production is overwhelmingly dominated by QCD interactions, an increase of the very small $V_{td}$ by a factor ${\cal O}(10)$ has no discernible effect on this background. For the backgrounds $\ell\ell j$ and $\ell\ell b$, the di-lepton pair arises from an intermediate $Z$ or $\gamma^*$, thus they are independent of $V_{td}$ at LO. The $t$-channel single top background on the other hand includes contributions sensitive to $V_{td}$ already at LO, but they are again very small for $|V_{td}|\lesssim{\cal O}(10^{-1})$. As an example: for $r=10$ ($r=20$) the production cross-section increases by $5\%$ ($20\%$). Therefore, for moderate values of $r$ this background does not have a significant growth. For $WWj$, there are Feynman diagrams depending on $V_{td}$, however, similar to the case of $t\bar t$, their contribution is tiny compared to the dominant contributions that are proportional to the diagonal elements of the CKM matrix. Thus, even for $r\sim{\cal O}(10)$ the impact on this background can be neglected. 

At last it should be mentioned that the partonic initial states $gs$ and $g\bar s$ can generate an irreducible background controlled by $V_{ts}$, namely $g s \to t W$.  However, this background has a cross-section suppressed by both the smallness of $|V_{ts}|$ and the strange quark PDF,  plus a negligible charge asymmetry, therefore we can safely neglect it even for values of $|V_{ts}|$ at the current experimental upper bound.

\subsection{Enhancing a charge asymmetric signal over a (mostly) symmetric background}

Given the above discussion we are left with a charge asymmetric signal in $pp \to t W$, and the approximately symmetric main backgrounds $pp\to t\bar t$ and $pp\to t W (b)$.   Other backgrounds are negligible or become negligible after a cut in $E_T^{\rm miss}$ and $m_{\ell\ell}$, such as $pp \to Z/\gamma^* j$.

\begin{figure}[t]
\centering
\includegraphics[width=0.48\textwidth]{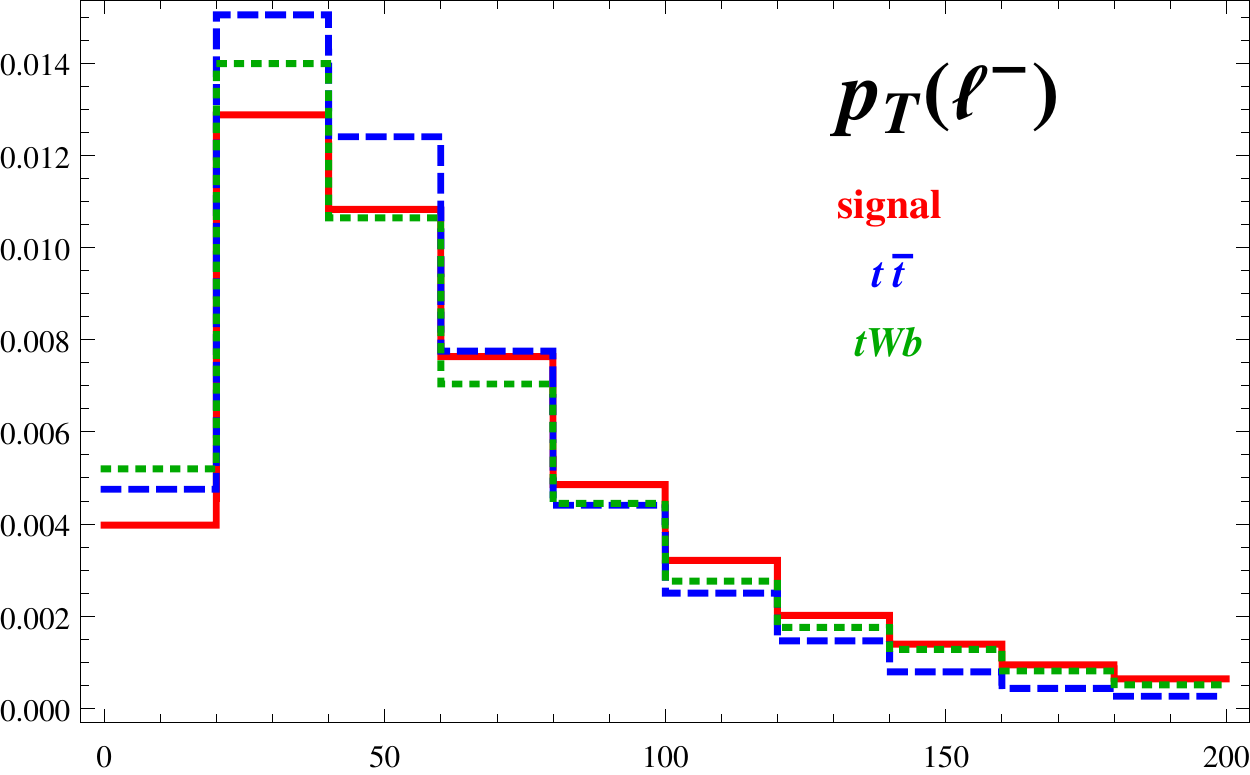}
\hspace{3mm}
\includegraphics[width=0.48\textwidth]{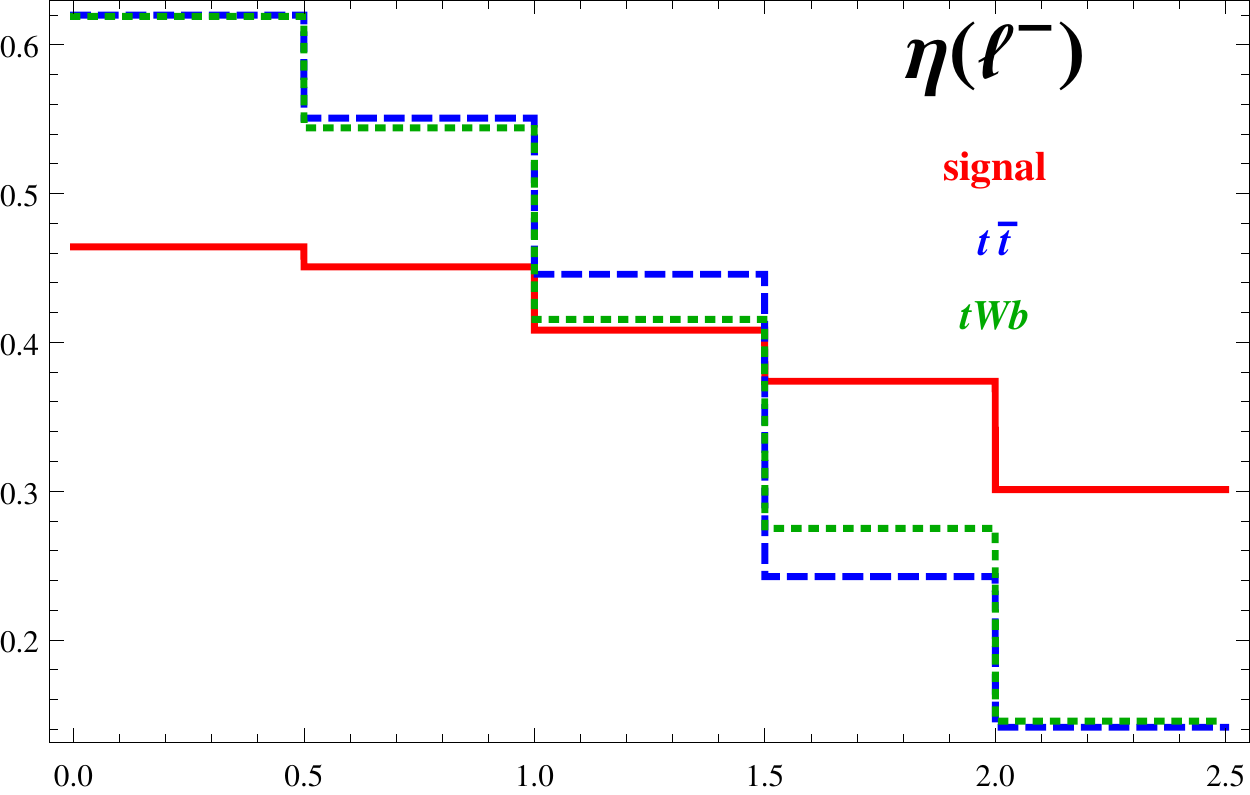}
\vskip0.5cm
\includegraphics[width=0.48\textwidth]{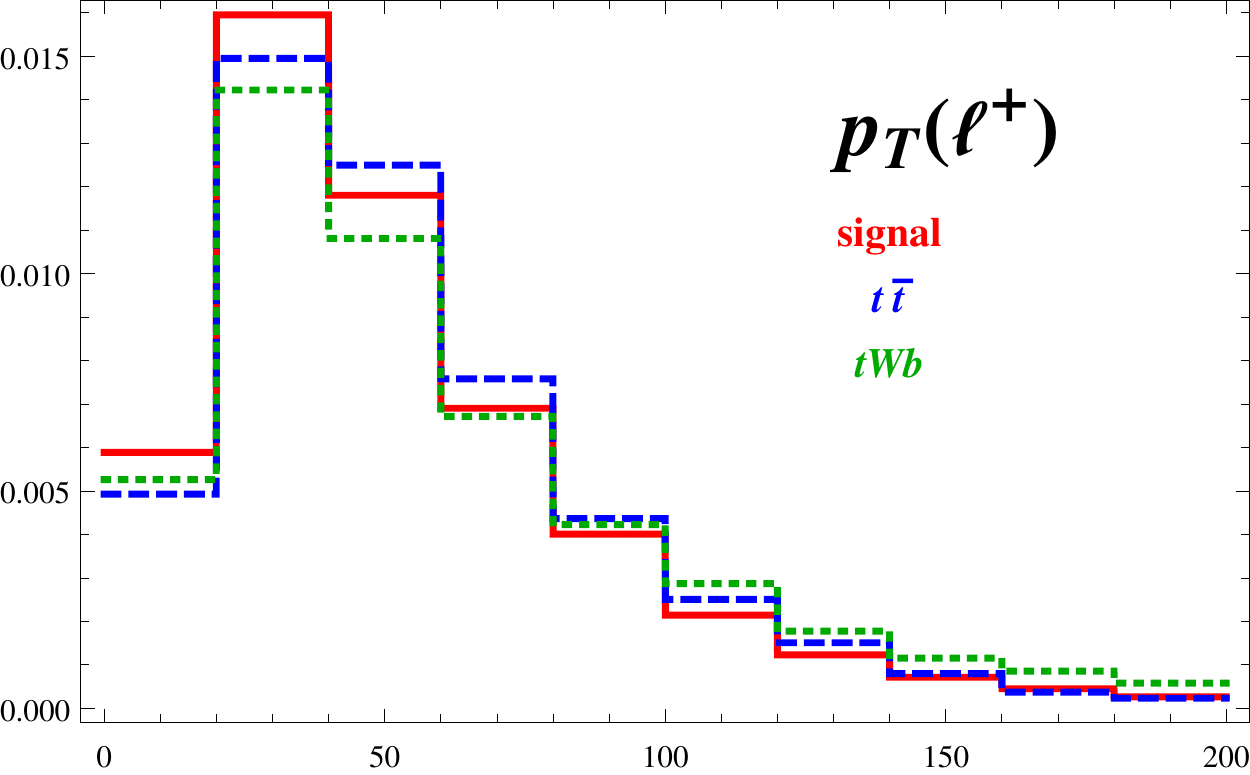}
\hspace{3mm}
\includegraphics[width=0.48\textwidth]{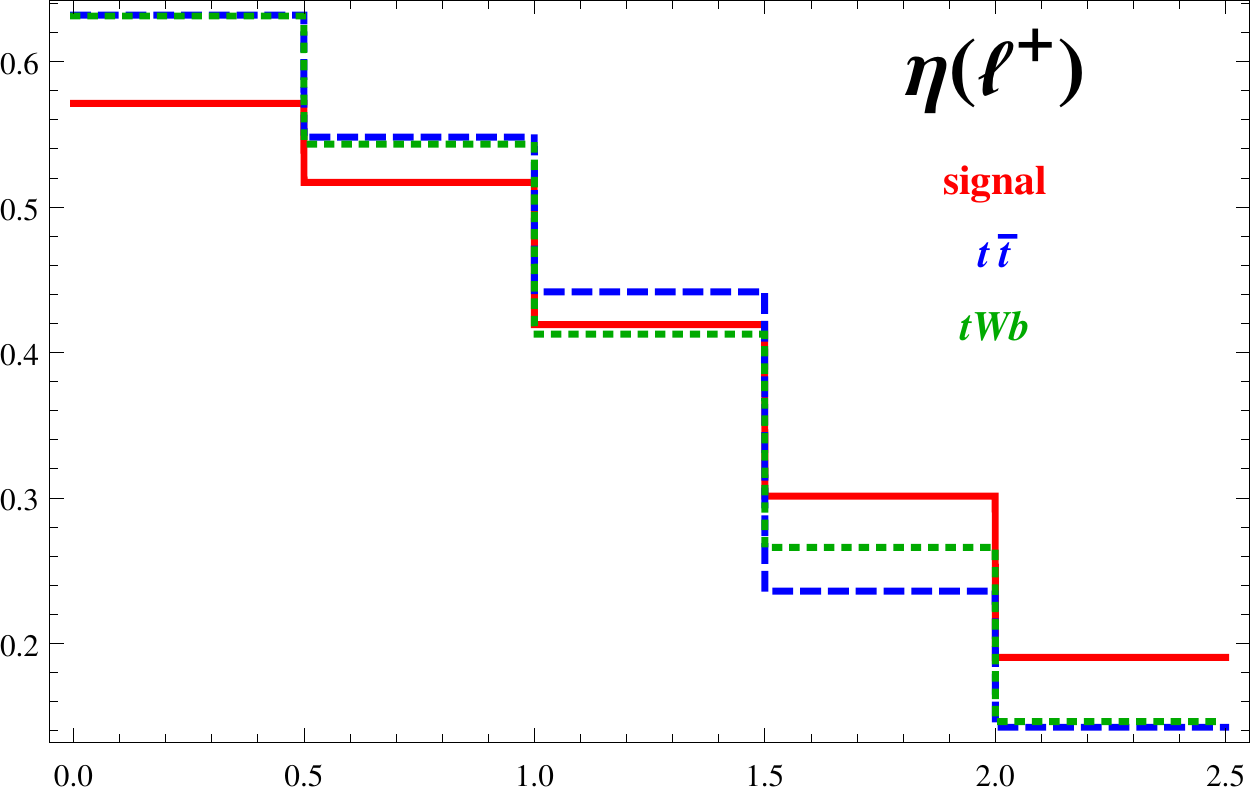}
\vskip0.5cm
\includegraphics[width=0.48\textwidth]{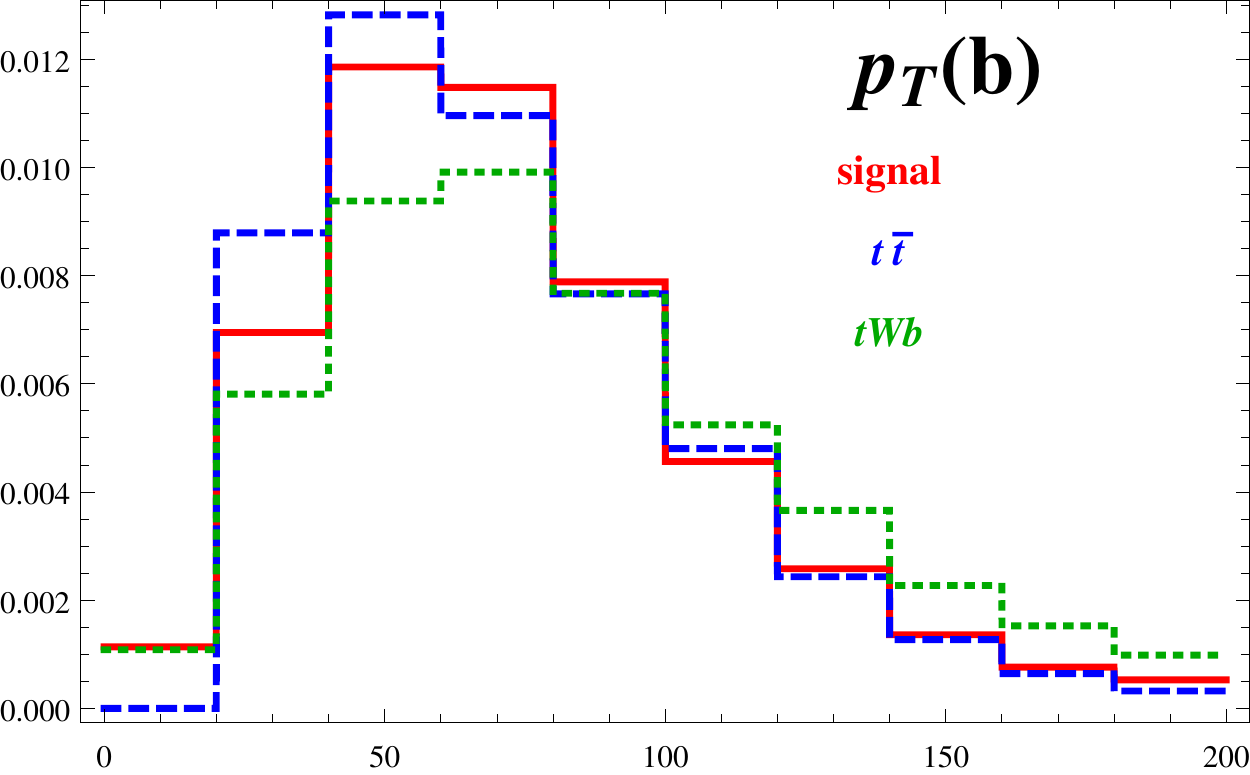}
\hspace{3mm}
\includegraphics[width=0.48\textwidth]{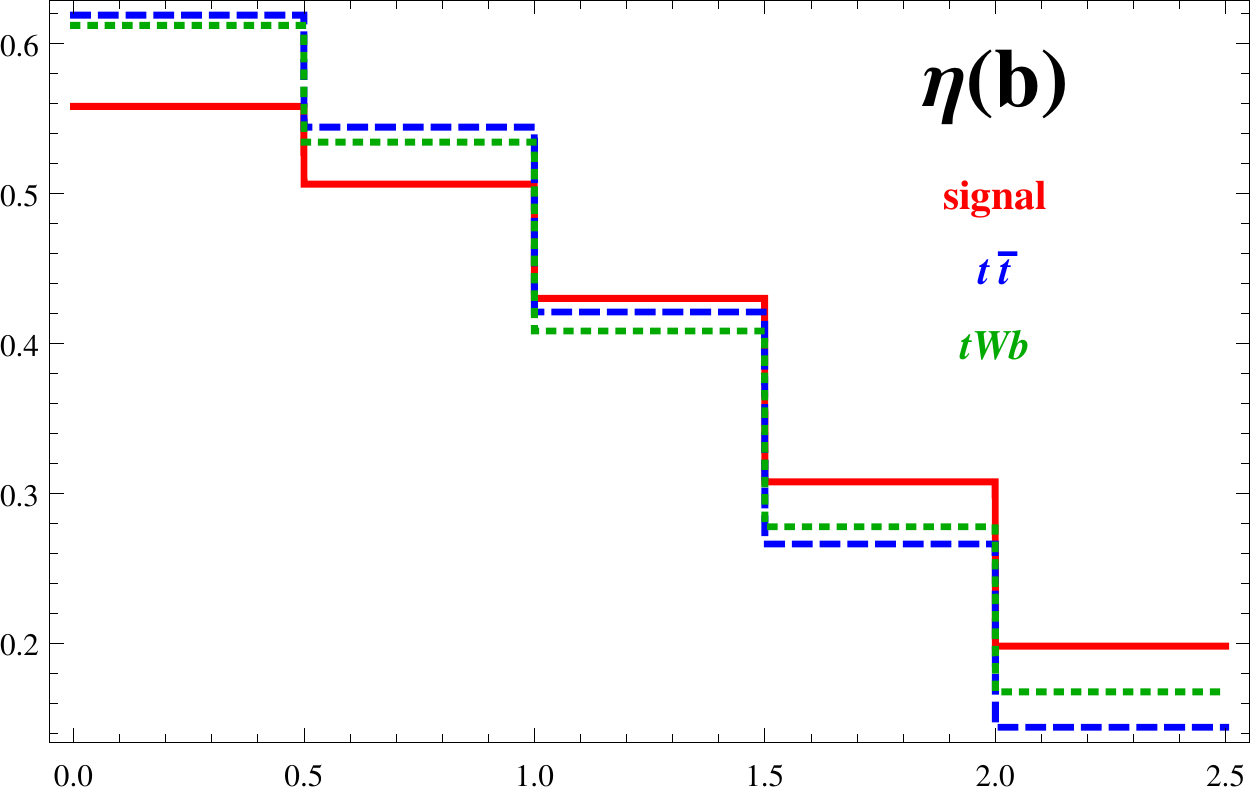}
\caption{Distributions of $p_T$ and $\eta$ of the final particles at parton level for the signal (red and continuous) and the main backgrounds $t\bar t$ (blue and dashed) and $tW(b)$ (green and dotted).} 
\label{distributions}
\end{figure}

In order to quantify the different signal and background features expected from the previous qualitative analysis, we first consider the relevant parton level distributions of the signal and the main backgrounds.  We have simulated the events at parton level using {\tt MadGraph5\_aMC@NLO} \cite{Alwall:2014hca}. The $tW(b)$ background has been simulated in the four-flavor PDF scheme, resulting in a $tWb$ final state.\footnote{See Ref.~\cite{Maltoni:2012pa} for discussion on the appropriate use of this scheme.} Thus, for both backgrounds we have required that only one of the $b$'s falls into the acceptance region defined by $|\eta(b_1)|<2.5$ and $p_T(b_1)>20$~GeV while the second $b$ is restricted to the regions $|\eta(b_2)|>5$ or $p_T(b_2)<20~GeV$, mimicking a jet-veto aimed predominantly at suppressing the $t\bar t$ background (more sophisticated methods for dealing with this overwhelming background are discussed in the next section).  The results are shown Fig.~\ref{distributions}.
We observe that the most important difference comes from the $\eta(\ell^-)$ distribution, where the signal clearly prefers forward negatively charged leptons, as expected from the qualitative discussion in Sect.~\ref{sec:Vtd}.  A similar preference is also present in the $\eta(\ell^+)$ and $\eta(b)$ distributions, although much less pronounced.  On the other hand, the $p_T$-distributions of the signal and the backgrounds do not offer as clear a differentiation as in the $\eta$ case.  For example, the $p_T(b)$ ($p_T(\ell^-)$) distributions could only be used to distinguish the signal from $tW(b)$ ($t\bar t$), respectively. 

\begin{figure}[t]
\centering
\includegraphics[width=0.3\textwidth]{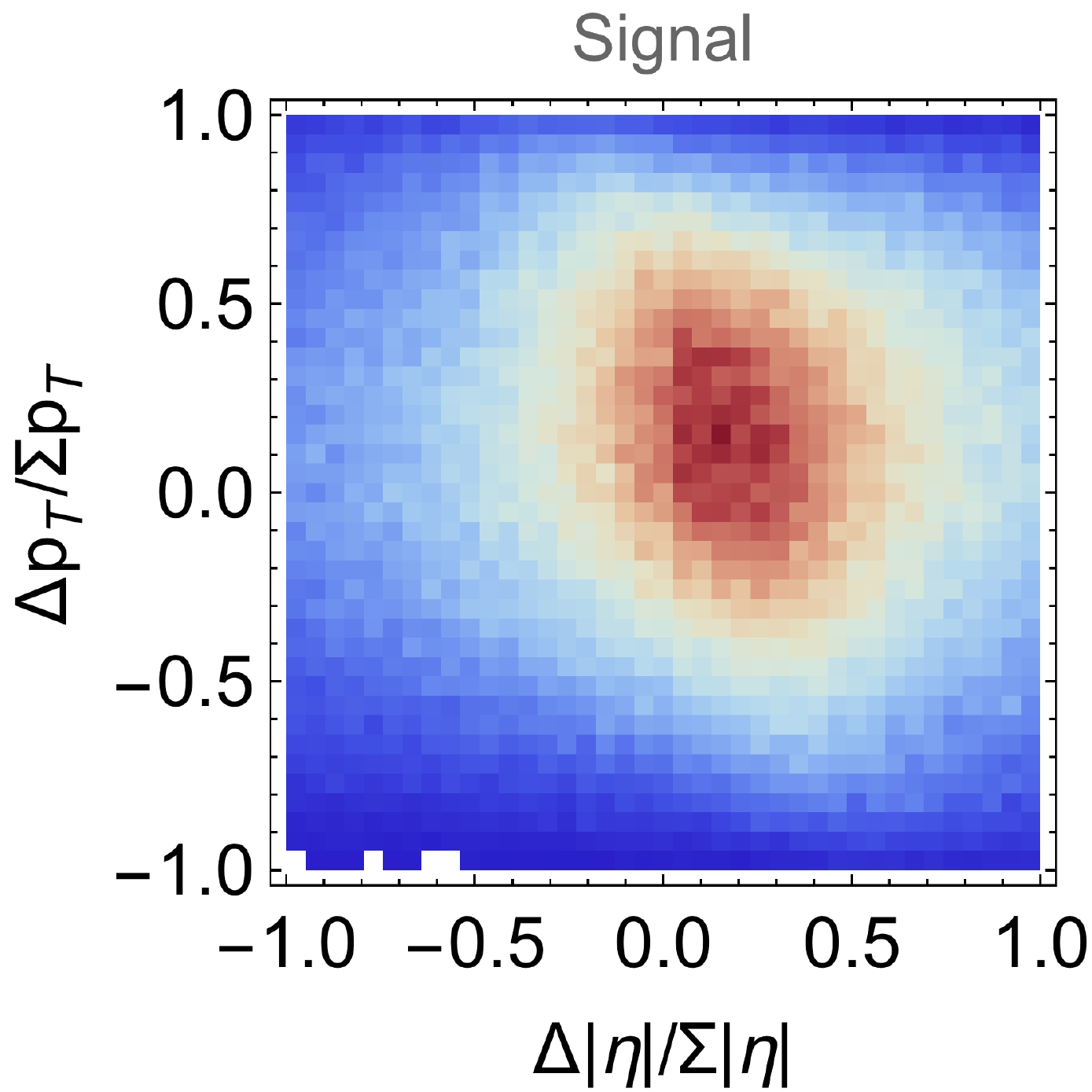}
\hspace{3mm}
\includegraphics[width=0.3\textwidth]{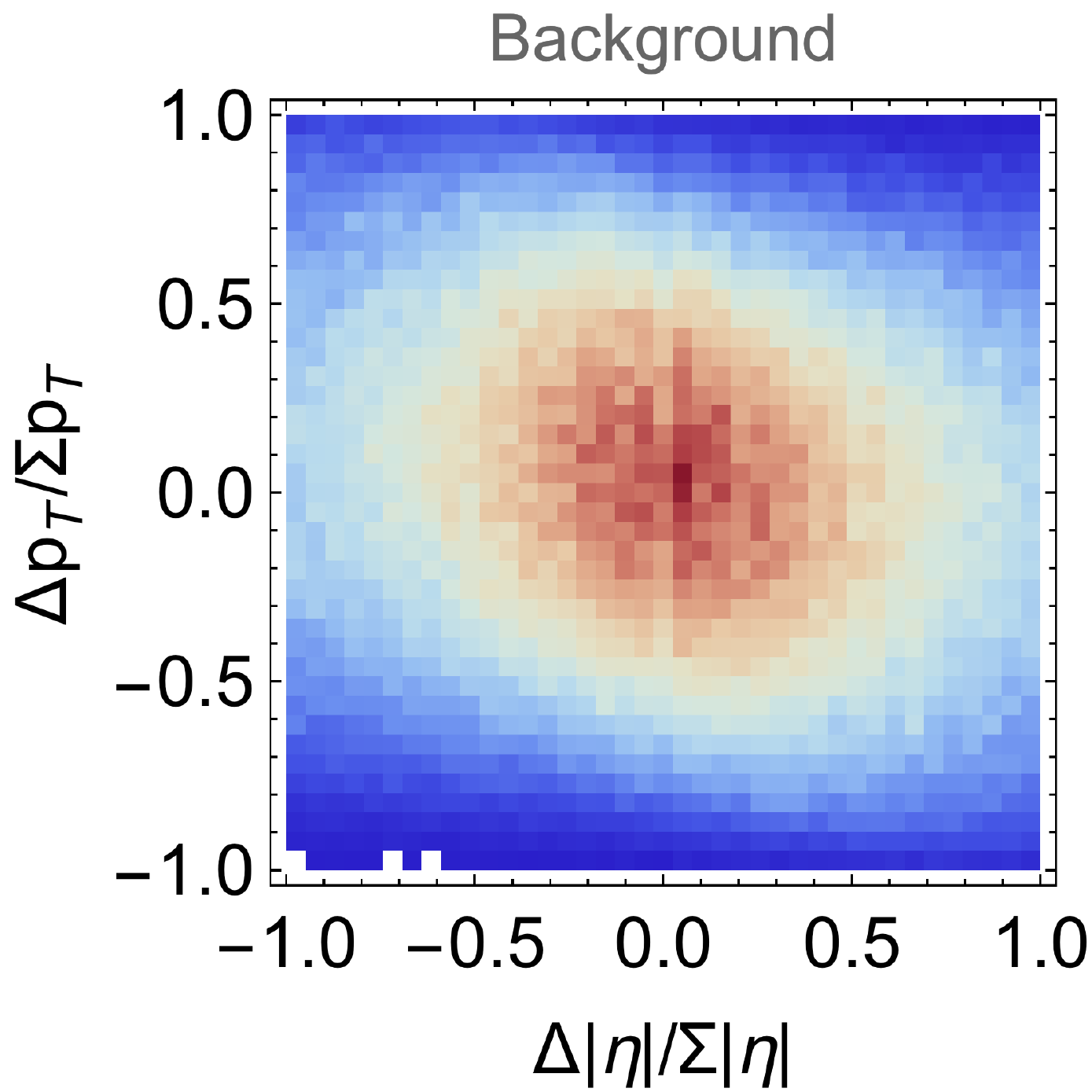}
\hspace{3mm}
\includegraphics[width=0.3\textwidth]{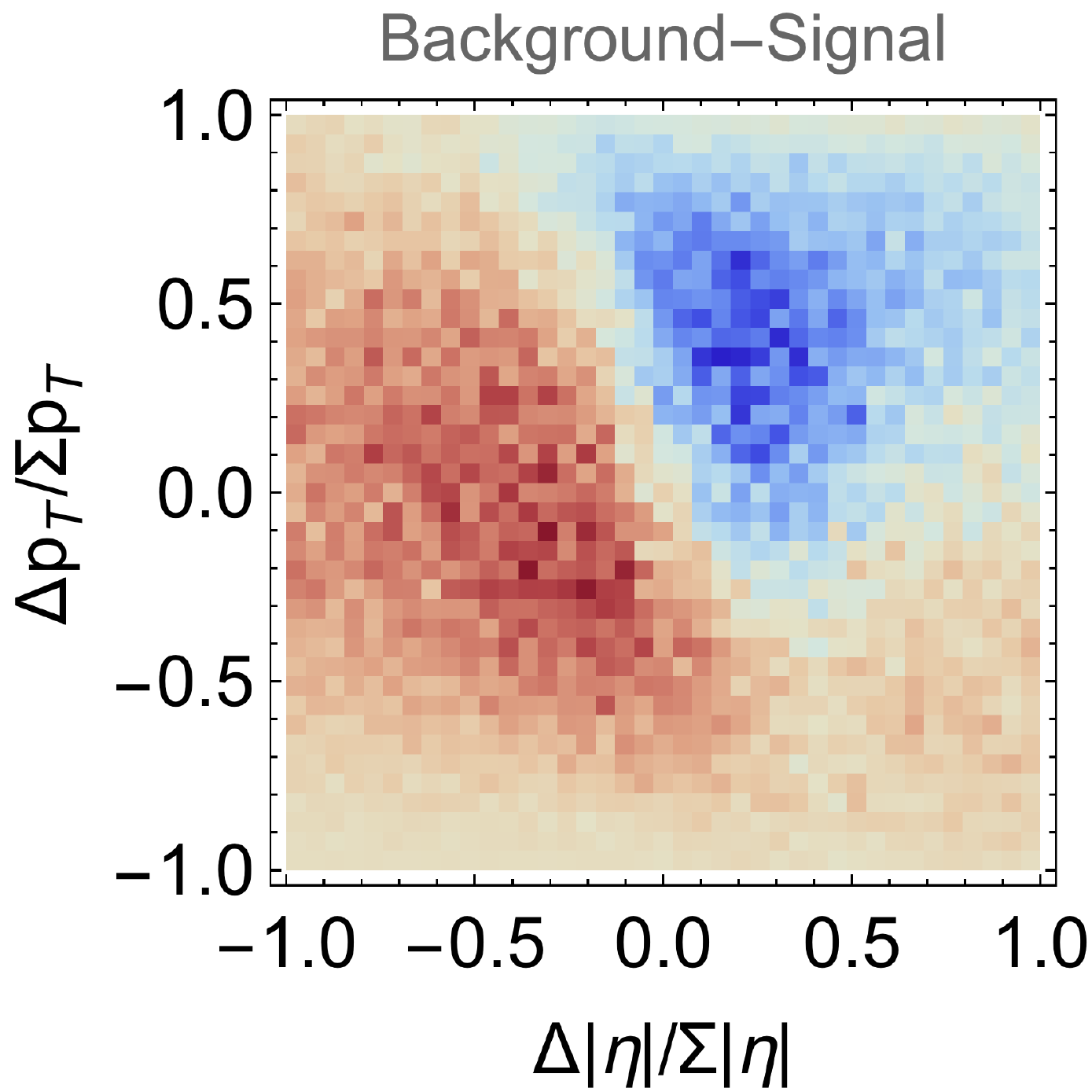}
\hspace{3mm}
\caption{Density plot of normalized event distributions of the signal (left) and the main background $t\bar t$ (center) in the $\Delta |\eta(\ell)|/\Sigma|\eta(\ell)|$ versus $\Delta p_T(\ell)/\Sigma p_T(\ell)$ plane.  In the right panel we show the difference between the two distributions, which demonstrates that (compared to the almost symmetric background) signal events are predominantly in the 3rd quadrant of the plot, whereas they are missing in the 1st quadrant.  This correlation between the plotted variables points out that an asymmetry between these quadrants would be useful to enhance signal over background.} 
\label{scatter}
\end{figure}

Given these distinctions between the signal and backgrounds in one-variable distributions, we next study distributions of pairs of variables in order to construct the most sensitive observables that could enhance the signal over the background.  With this purpose, and motivated by the results in Fig.~\ref{distributions}, we plot in Fig.~\ref{scatter} the simultaneous (normalized) distributions of the signal and the main ($t\bar t$) background in the $\Delta |\eta(\ell)|/\Sigma |\eta(\ell)|$ -- $\Delta p_T(\ell)/\Sigma  p_T(\ell)$ plane, where
\begin{align}
\Delta |\eta(\ell)| &= |\eta(\ell^+)| - |\eta(\ell^-)| \,, & \Sigma |\eta(\ell)| &= |\eta(\ell^+)| + |\eta(\ell^-)| \,,\nonumber \\
\Delta p_T(\ell) &= p_T(\ell^+) - p_T(\ell^-)  \,, & \Sigma p_T(\ell) &= p_T(\ell^+) + p_T(\ell^-) \,.
\end{align}
As it can be seen in the figure, a sizable asymmetry in the signal can be constructed by comparing events in the first and third quadrants.  
Thus we propose the following asymmetry
\begin{equation}
A(\eta,p_T) = \frac{N^+ - N^-}{N^+ + N^-},
\label{A}
\end{equation}
where
\begin{equation}
N^\pm = N\left( \Delta |\eta(\ell)| \gtrless 0 \,\& \, \Delta p_T(\ell) \gtrless 0 \right),
\label{i-th contribution}
\end{equation}
as a $|V_{td}|$ sensitive observable.

It is interesting to understand how the different processes contribute in $A(\eta,p_T)$.  It is clear that the denominator is dominated by $t\bar t$, whose cross-section is considerably larger than the others, even after the selection cuts.  On the other hand, the numerator is more involved because each $i$-th process contributes with 
\begin{equation}
N^+_i - N^-_i = \sigma_i \cdot {\cal A}_i \cdot \epsilon_i \cdot A_i(\eta,p_T),
\end{equation}
where the factors in the RHS are the cross-section, acceptance, selection efficiency\footnote{Selection efficiency refers to the fraction of events that pass the selection cuts to be either $N^\pm$.} and asymmetry, respectively, all restricted exclusively to the $i$-th process.  Since $t\bar t$ has a small NLO charge-asymmetry, but a large cross-section, it ends up being important for the $r\approx 1$ region, but becomes sub-leading as $r \gtrsim 10$.  We expect a similar NLO asymmetry for $tW(b)$, but since its cross-section is much smaller than $t\bar t$ we can neglect it in the numerator. Other backgrounds such as $WWj$, $tj$ and  $Z/\gamma^* j$ have a non-negligible asymmetry, but their product $\sigma_i \cdot {\cal A}_i \cdot \epsilon_i $ suppresses any contribution to the numerator. 

\subsection{Other asymmetries}

Given the distributions in Fig.~\ref{distributions} one could consider alternative definitions of asymmetries to enhance the signal over the backgrounds. We have tested many of them bearing in mind that we need to exploit the lepton charge asymmetry present in the signal. In the following paragraphs, we explain the main ones and why they do not improve the significance of the asymmetry defined in Eq.~(\ref{A}).

The most interesting attempt consists of taking the asymmetry $A(\eta,p_T)$ with a cut in $|\eta_{\ell^-}|\gtrsim 1.5$ since we expect to have an enhanced asymmetry in the forward region. Indeed this selection increases the absolute value of the asymmetry, but reduces the acceptance due to the additional cut. Moreover, this also creates an artificial asymmetry in the backgrounds which reduces the sensitivity to the signal. Together, this results in larger statistical (and also presumed systematic) uncertainties and consequently reduced significance when compared to $A(\eta,p_T)$. On the other hand, imposing a symmetric cut on $|\eta_{\ell^-}|$ and $|\eta_{\ell^+}|$ keeps the backgrounds symmetric but does not increase the signal asymmetry sufficiently to offset the reduction in the acceptance.

The last case we discuss is the asymmetry based solely on $|\Delta \eta|$, with and without cuts on $\eta$, that is
\begin{equation}
A( \eta) = \frac{N\left( \Delta |\eta(\ell)| > 0\right)  - N\left( \Delta |\eta(\ell)| < 0 \right)}{N\left( \Delta |\eta(\ell)| > 0 \right) + N\left( \Delta |\eta(\ell)| < 0 \right)} .
\end{equation}
In this case the total number of accepted events increases and therefore the statistical uncertainty decreases.  However, the absolute value of the asymmetry decreases because  $\Delta |\eta|$ alone has less discriminating power than $\Delta |\eta|$ and $\Delta p_T$ together (see right panel in Fig.~\ref{scatter}).  The combination of these two features yields smaller significance than $A(\eta, p_T)$. While this asymmetry is also sensitive to cuts in $\eta$, we have found that their application does not improve the significance.

%
\section{Results}
\label{sec:results}
%

For our final quantitative analysis and estimation of the experimental reach, we have simulated the signal and the main backgrounds using {\tt MadGraph5\_aMC@NLO}~\cite{Alwall:2014hca}, interfaced with {\tt Herwig} \cite{Bellm:2015jjp,Bahr:2008pv} (for $t\bar t$) and {\tt Pythia8} \cite{Sjostrand:2014zea,Sjostrand:2006za} (for all other processes) for showering and hadronization. The $t\bar t$ has been simulated at NLO in QCD to account for its non-vanishing charge asymmetry.  The NLO effects in the other relevant processes have been accounted for through the effective $k$-factors. In particular, $k_{tW(b)}=1.35$~\cite{Kidonakis:2007wg} and we assume the same $k$-factor for the signal. Finally, we have simulated detector effects using {\tt Delphes} \cite{deFavereau:2013fsa}. The jets have been clustered using the anti-kt algorithm with $R=0.6$. A `loose' $b$-tagging algorithm working point has been used with a reference selection efficiency of $0.8$ and a rejection factor for light jets of $100$ \cite{ATL-PHYS-PUB-2015-022,deFavereau:2013fsa}. 
The remaining Delphes parameters have been left in the default ATLAS tune.

Since the signal results in a $\ell^+ \ell^- b\, E_T^{\rm{miss}}$ final state, we have suppressed the main backgrounds through the following selection of events:
\begin{itemize}
\item Select events with $\ell^+ \ell^- b$, all with $|\eta|<2.5$ and $p_T>20$ GeV.
\item Veto events with additional jets within $|\eta|<5$. (Suppress $t\bar t$.) 
\item Veto events with $E_T^{\rm{miss}} < 30$ GeV or $m_{\ell\ell}<25$ GeV or $|m_{\ell\ell}-m_Z|<15$ GeV. (Suppress $Z/\gamma^* j$.)
\end{itemize}
We have checked the sensitivity of the selection cuts on the event reconstruction parameters and found that the jet veto depends quite sensitively on the jet clustering algorithm. In particular, it becomes less efficient for narrower jets.     

\begin{figure}[t]
\centering
\includegraphics[width=0.7\textwidth]{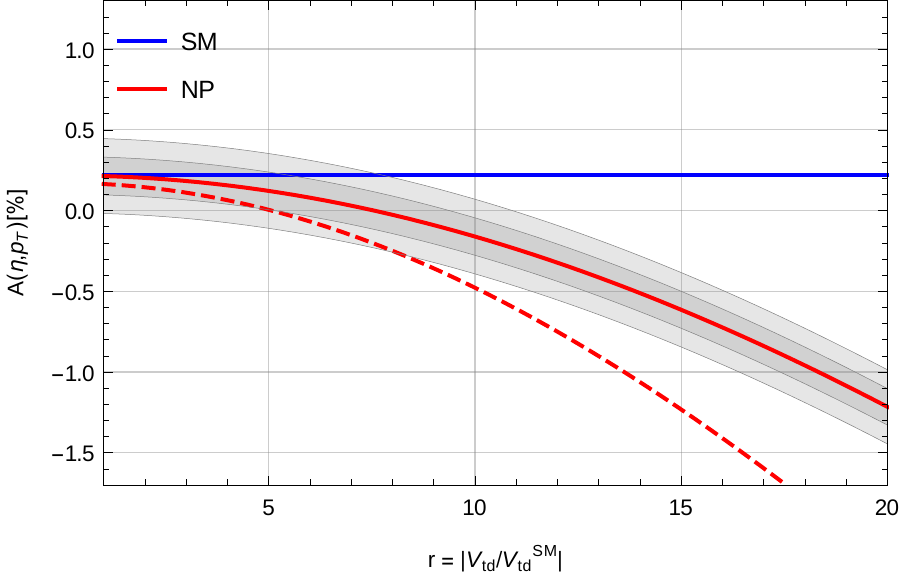}
\caption{Expected value of the asymmetry $A(\eta,p_T)$ (solid red) as function of the NP parameter $r= |V_{td}/V_{td}^{\rm SM}|$. Shaded area represents statistical (darker) + assumed systematic (lighter) uncertainties ($\pm1\sigma$) with the event selection indicated in text at the prospective LHC luminosity of $L=3000$ fb$^{-1}$.  Blue line represents the SM value of the asymmetry, which is mainly due to NLO QCD effects in $t\bar t$.  Also shown (in dashed red) is an estimation of the asymmetry assuming a reduction of the dominant $t\bar t$ background by half (see text for details).}
\label{Aplot}
\end{figure}

In Table \ref{flow} we show how the signal and the main backgrounds behave upon detector effects and selection cuts for the above described event reconstruction and selection.  Using  the fifth and the last column one can visualize the importance of each contribution to the denominator and the numerator of the asymmetry defined in Eq.~\eqref{A}, respectively.
\begin{table}
\begin{center}
  \begin{tabular}{ | L{2cm} || C{2cm} | C{2.5cm} | C{1.8cm} | C{2.5cm} | C{2.5cm} | C{2.8cm} |}
    \hline
    process & $\sigma \cdot \mathcal B$ [fb]& ${\cal A} (\ell^+\ell^-b + X)$ & $\epsilon$ & $\sigma \cdot  \mathcal B \cdot {\cal A} \cdot \epsilon$ [fb] & $A_i(|\Delta \eta|,\Delta p_T)$ & $\sigma \cdot  \mathcal B \cdot {\cal A} \cdot \epsilon\cdot A_i$ [fb]  \\ \hline \hline
    signal & 1.2$\, r^2$  & 0.17 & 0.16 & 0.034$\, r^2$ & -0.2& -0.0067$\,r^2$\\  \hline
    $t\bar t$ & $3.1 \times 10^5$   & 0.56 & 0.011 & 200 & 0.003 & 0.57 \\ \hline
    $tW(b)$ & $1.8 \times 10^3$  & 0.25 & 0.07 & 34 & ${\cal O}(10^{-3})$ & ${\cal O}(10^{-2})$\\  \hline
    $Z/\gamma^* j$ & $5.1 \times 10^5$ & 0.002 & $4.7\times 10^{-4}$ & 0.53 & -0.10 & -0.05 \\  \hline
    $WWj$ & $1.5\times 10^3$ & 0.002 & 0.14 & 0.52 & -0.06 & -0.03 \\  \hline
    $tj$ & $1.7 \times 10^4$ & $1.2\times 10^{-5}$ & 0.29 & 0.0062 & -0.8 & -0.02 \\  \hline
  \end{tabular}
\end{center}
\caption{The relevant processes upon detector acceptance and selection cuts. ${\cal A}$ refers to the detector acceptance of $\ell^+\ell^-b$ plus anything else.  Selection efficiency $\epsilon$ includes a veto on events with extra jets, cuts in $E_T^{\rm{miss}}$ and $m_{\ell\ell}$, and also a selection of events only in the first and third quadrants in $\Delta |\eta|$ and $\Delta p_T$, as defined in Eq.~\eqref{i-th contribution}.  
The column `$\sigma \cdot \mathcal B \cdot {\cal A} \cdot \epsilon$' is relevant for the denominator of the total asymmetry and is dominated by $t\bar t$ and a small correction by $tW(b)$.  
$A_i$ refers to the asymmetry defined in Eq.~\eqref{A} for the corresponding row, constructed with the detector level events that pass all acceptance and selection requirements. 
Finally, the last column is relevant for the numerator of the total asymmetry and is dominated by $t\bar t$ for $r\lesssim10$ and by the signal for $r\gtrsim10$.}
\label{flow}
\end{table}
In Fig.~\ref{Aplot} we plot the resulting charge asymmetry $A(\eta,p_T)$ defined in Eq.~(\ref{A}) with all the detector level simulations included.  One can see that the $t\bar t$ asymmetry dominates for $r=\mathcal O(1)$ (close to the SM), but as $r$ increases the negative contribution from the signal starts to dominate.

\begin{figure}[t]
\centering
\includegraphics[width=0.8\textwidth]{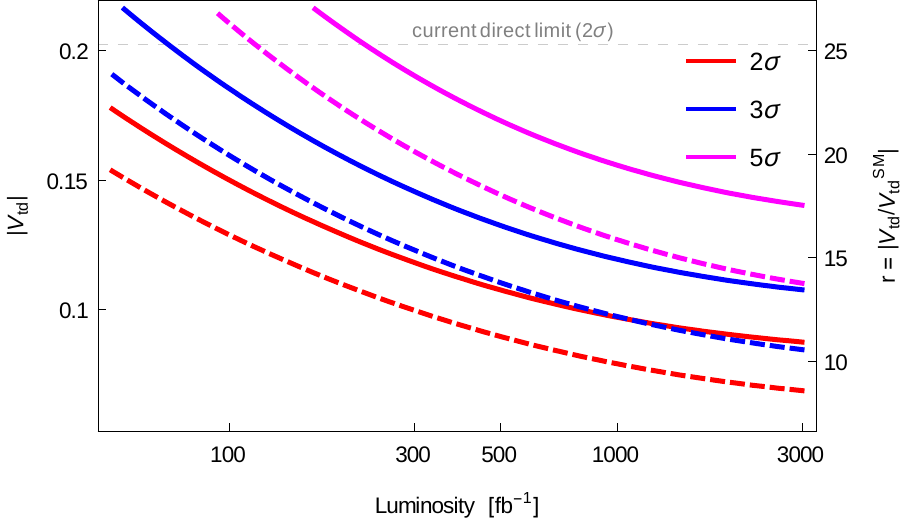}
\caption{Contour lines for the $2\sigma$, $3\sigma$ and $5\sigma$ measurement of $|V_{td}|$ parameterized as a function of $r=|V_{td}/V_{td}^{\rm SM}|$ and the LHC luminosity.  Dashed lines correspond to the estimation of the same analysis assuming a further reduction of the $t\bar t$ background by half (see text for details).  }
\label{significance}
\end{figure}

To quantify the versatility of the proposed charge asymmetry we have studied the prospective experimental reach in the NP parameter $r$ by computing the difference of $A(\eta,p_T)$ to its SM expectation in units of the uncertainty.  In addition to the statistical uncertainty we have included an estimation for the systematic uncertainty $\Delta_{\rm syst} = 0.2 \,\% $\,, based on a similar analysis in the di-lepton charge asymmetry performed by CMS in Ref.~\cite{Naseri:2017gzt} {and the expected usual improvement in the knowledge of the detector and other systematic effects with increasing luminosity}.  In our analysis $\Delta_{\rm syst}$ acts as an overall estimation of all the systematic uncertainties.  By adding statistical and systematic uncertainties in quadrature, we define the significance as
\begin{equation}
\mbox{significance} = \frac{\left| A(\eta,p_T) - A(\eta,p_T)^{\rm SM}\right| }{\sqrt{(N^+ + N^-)^{-1} + \Delta_{\rm syst}^2}} .
\end{equation}
In Fig.~\ref{significance} we plot contours of expected significance in measuring $A(\eta,p_T)$ as a function of $r$ and the luminosity.  In order to further differentiate between the small signal and  the overwhelming $t\bar t$ background, existing experimental analyses \cite{Aaboud:2016lpj,Aad:2015eto,Chatrchyan:2014tua} of $tW$ associated production at the LHC, in addition to basic selection cuts similar to the ones described above, employ more elaborate multivariate techniques, such as boosted decision trees or neural networks. With the rapid development of machine learning, these methods are expected to be further refined in the near future and also extremely useful for the processes and observables studied here.  As a rough estimation, and motivated by Ref.~\cite{Chatrchyan:2014tua}, we include in Fig.~\ref{significance} (in red dashed line) an estimation of the significance for the case where the $t\bar t$ background could be reduced by a further factor of $1/2$ with negligible effect on the signal.

We observe that in using the proposed charge asymmetry $A(\eta,p_T)$ in the leptonic $tW$ final state, an improvement in the direct bound on $|V_{td}|$ ($r$) compared to existing constraints is possible already with the existing LHC dataset. Furthermore,  values of $r<10$ could be directly accessible at the (HL)LHC, improving the existing best direct constraints by roughly a factor of three. A further significant improvement would however require a reduction of systematic uncertainties below the per-mille level.

%
\section{Conclusions}
\label{sec:conclusions}
%
The CKM elements $V_{td}$ and $V_{ts}$ are fundamental parameters of the SM governing flavor conversion in the top sector. Their determination through direct measurements is a difficult task that requires processes with on-shell top quarks. We have proposed an observable that can test $|V_{td}|$ to ${\cal O}(10^{-1})$ in the creation of $tW$  at the LHC. Selecting a final state with $\ell^+\ell^- b$, we have defined a charge asymmetry using the variables $\eta$ and $p_T$ of the leptons, that is sensitive to $|V_{td}|$. We have studied and characterized the signal and main backgrounds at parton level, as well as by using simulations up to (parametric) detector level. We have shown that, although the backgrounds have overwhelming production cross sections, they are highly symmetric and an asymmetric signal can eventually emerge. We have computed the asymmetry as a function of $|V_{td}|$, and determined the prospective reach of the LHC as a function of the luminosity. We have shown that the current bound on direct determination of $|V_{td}|$ can be surpassed with the existing LHC dataset, and that with $3000{\rm fb}^{-1}$ it could be possible to exclude $|V_{td}| \gtrsim 0.1$ at the $2\sigma$ level.

The dominant $t\bar t$ background, although being charge-symmetric at leading order, strongly suppresses the asymmetry by giving a large contribution to its denominator. A crucial improvement upon our cut-based approach would therefore be to further reduce this background while preserving the signal (using e.g.~multivariate or machine-learning techniques). As an example, we have shown in Fig.~\ref{significance} that, by lowering $t\bar t$ by a factor 2, it would be possible to exclude $|V_{td}|\gtrsim 0.1$ already with $600{\rm fb}^{-1}$ of luminosity. Finally, a further significant reduction in systematic uncertainties below our current estimate of $0.2\%$ could allow the high luminosity LHC eventually to probe values as low as $|V_{td}| \sim 0.06$\,. 

We have also studied a number of alternative definitions of the charge asymmetry, including asymmetries only in $\eta(\ell)$, as well as the implementation of additional cuts that could increase their size. We found that in all the cases the total uncertainty increases, and the significance is at best comparable with the original one.

Finally, we comment on other processes that could also give valuable information for the direct determination of $V_{tq}$ at the LHC. For example, kinematical distributions in t-channel single top production could also be used to probe $V_{ts}$ (and $V_{td}$) suppressed contributions~\cite{Lacker:2012ek}. A much less explored example is however $pp\to W^+W^-$ that is sensitive to $V_{td}$ and $V_{ts}$ and can be studied at the LHC. It can be complementary to the observables proposed in this paper and in the existing literature and certainly deserves a detailed study. 

\begin{acknowledgments}
JFK acknowledges the financial support from the Slovenian Research Agency (research core funding No. P1-0035 and J1-8137).
This work was partially supported by the Argentinian ANPCyT PICT 2013-2266 and by the cooperation agreement MHEST-MINCyT  SLO-14-01, ARRS BI-AR/15-17-001, between Slovenia and Argentina.
\end{acknowledgments}

\bibliography{biblio}

\end{document}